\documentclass[letter,usenatbib]{aa} 
\usepackage[usenames]{color}

\usepackage{graphicx}
\usepackage{txfonts}
\usepackage{epstopdf}
\usepackage[usenames]{color}

\newcommand{\logg}{ {\rm log\,{\it g}}}
\newcommand{\teff}{ T_{{\rm eff}}}
\newcommand{\feh}{{\rm [Fe/H]}}
\newcommand{\lum}{{\rm log}\, L/L_{\odot}}

\newcommand{\thetaLD}{\theta_\mathrm{LD}}

\usepackage{natbib}
\bibpunct{(}{)}{;}{a}{}{,}
\begin{document}
   \title{The diameter of the CoRoT target HD 49933}

   \subtitle{Combining the 3D limb darkening, asteroseismology, and interferometry}

   \author{L. Bigot\inst{1}\mail{lbigot@oca.eu},  D. Mourard\inst{2}, P.
Berio\inst{2}, F.
Th\'{e}venin\inst{1}, R. Ligi\inst{2}, I. Tallon-Bosc\inst{3}, O.
Chesneau\inst{2}, O. Delaa\inst{2}, N. Nardetto\inst{2}, K. Perraut\inst{4}, Ph.
Stee\inst{2},  T. Boyajian\inst{5}, P. Morel\inst{1}, B. Pichon\inst{1}, P.
Kervella\inst{6}, F.~X. Schmider\inst{2}, H. McAlister\inst{7,8}, T.
ten~Brummelaar\inst{8}, S.~T. Ridgway\inst{9}, J. Sturmann\inst{8}, L.
Sturmann\inst{8}, N.
Turner\inst{8}, C. Farrington\inst{8} and P.~J. Goldfinger\inst{8}}
   \institute{Universit\'e Nice-Sophia Antipolis, Observatoire de la C\^{o}te
d'Azur,  CNRS UMR 6202, BP 4229, F-06304 Nice Cedex, France.
\and Universit\'e Nice-Sophia Antipolis, Observatoire de la
C\^{o}te d'Azur,  CNRS UMR 6525, BP 4229, F-06304 Nice Cedex, France.
\and UCBL/CNRS CRAL, 9 avenue Charles Andr\'{e}, 69561 Saint Genis Laval cedex, France
\and UJF-Grenoble 1 / CNRS-INSU, Institut de Plan\'etologie et d'Astrophysique de
Grenoble, UMR 5274, Grenoble, F-38041, France.
\and CHARA and Department of Physics and
Astronomy, Georgia State University, P.O. Box 4106, Atlanta, GA 30302-4106, USA
\and LESIA, Observatoire de Paris, CNRS UMR 8109, UPMC, Universit\'e Paris Diderot,
5 place Jules Janssen, 92195, Meudon, France
\and Georgia State University, P.O. Box 3969, Atlanta GA 30302-3969, USA
\and CHARA Array, Mount Wilson Observatory, 91023 Mount Wilson CA, USA
\and National Optical Astronomy Observatory, P.~O. Box 26732, Tucson, AZ 85726, USA
}

   \date{Received ; accepted }


  \abstract
   {
 The interpretation of stellar pulsations in terms of internal structure depends on
the knowledge of the fundamental stellar parameters. 
Long-base interferometers permit us to determine very accurate stellar radii, which are
independent constraints for stellar models that help us to locate the star in the HR
diagram.}
   {Using a direct interferometric determination of the angular diameter and
advanced three-dimensional (3D) modeling, we derive the radius of the CoRoT target HD49933 and reduce
the global stellar parameter space compatible with seismic data.}
   {The VEGA/CHARA spectro-interferometer is used to measure the angular diameter of
the star.  A 3D radiative hydrodynamical simulation of the surface is performed to compute the limb darkening and derive a reliable diameter from
visibility curves. The other fundamental stellar parameters (mass, age, and $\teff$) are found by fitting the large and small p-mode frequency separations
using a stellar evolution model that includes microscopic diffusion.}
   {We obtain a limb-darkened angular diameter of $\thetaLD = 0.445 \pm 0.012$ mas.
With the Hipparcos parallax, we obtain a radius of ${\rm R=1.42\pm
0.04}$ ${\rm R_{\odot}}$. The corresponding stellar evolution model that fits
both large and small frequency separations has a mass of $1.20\pm 0.08$ ${\rm
M_{\odot}}$ and an age of $2.7$ Gy. The atmospheric parameters are 
$\teff$ = 6640 $\pm \,100\,${\rm K}, \logg $= 4.21\pm 0.14$, and ${\rm 
[Fe/H]=-0.38}$.}
   {}


\titlerunning{The diameter of the CoRoT target HD49933}
\authorrunning{Bigot et al.}
   \maketitle

%
\begin{figure*}[ht!]
\includegraphics[height=31.5mm,width=180mm]{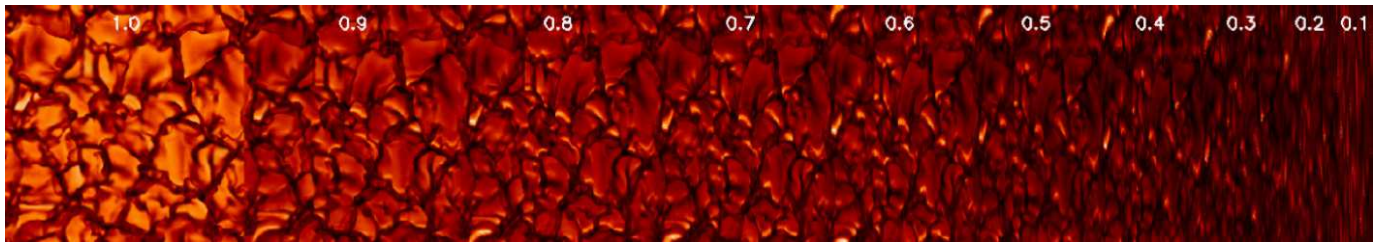}
\caption{Monochromatic  ($730$ nm)  center-to-limb emergent intensity at a
representative time of the simulation for various cosines of the viewing angle :
$\mu=1.0$ (disk center) down to $0.1$ (towards the limb). The horizontal sizes of
the simulation domain are $21 000\times 21000$ km.}
\label{granulation}
\end{figure*}
\section{Introduction}

The F5V solar-like pulsator HD 49933 (HR 2530, HIP 32851) is one of the primary
targets of the CoRoT space mission \citep{baglin06}.
The discovery of its oscillation was made from the ground by observing highly
resolved time-series spectra \citep{mosser05}. Since the launch of the spacecraft in
December 2006, two runs of 60 and 137 days have been performed leading to a rich
p-mode spectrum of 51 frequencies corresponding to the $\ell=0, 1, 2 $ eigenmodes
\citep{appourchaux08, benomar09, kallinger10}.

The interpretation of these eigenmodes to derive constraints on the stellar interior 
depends on the accuracy of the determination
of the fundamental parameters of the star ($\teff, \logg,\feh$). All chemical
abundance analyses of HD 49933 converge to that of a slightly metal-poor star with ${\rm
[Fe/H}]$ between -0.3 and -0.5 \citep[e.g.][]{gillon06, bruntt09}. The determination
of its effective temperature and gravity have been revisited several times. As
reported in \cite{bruntt09}, the $\teff$ obtained by different methods ranges from
$6450$ to $6780$ K,  which is by far too uncertain to tightly constrain the stellar
evolution models in the HR diagram. Similarly, $\logg$ is also not very well
constrained with values ranging from $4.0$ to $4.3$.
Several authors have attempted to determine $\teff$ and $\logg$ by fitting the ${\rm
H_\beta}$ and Mg {\scriptsize I} lines, respectively \citep{ryab09, kallinger10}.
Their solutions are most consistent with a cooler temperature ($\approx 6500$
K) and a lower gravity ($\approx 4.0$). \\ 
Since spectrometry and photometry lead to stellar parameters with a large
uncertainty, in this work we use a different approach by determining its diameter
using interferometry. Combined with seismology, it provides a very good constraint
of the stellar parameter space \citep{creevey07}. This approach was proposed
for the first time by \cite{kervella03} for the binary star $\alpha$Cen~A$\&$B and
later with similar success for other asteroseismic targets 
\citep{cunha07,teixeira09,bruntt09b,mazumdar09,bazot11}.

In this letter  we present the first campaign of interferometric measurements of HD
49933 obtained using the VEGA
instrument \citep{vega,vega2} at the CHARA Array \citep{chara}. 
We used state-of-art 3D hydrodynamical simulations of the surface of the star to
calculate
the limb darkening and to compute the visibility curves prior extracting the
angular diameter. The derived stellar radius is then used to constrain the other
fundamental stellar parameters by fitting the oscillation frequency separations.


\section{CHARA/VEGA observations}
Observations of HD 49933 were performed on October 16, 2010 using the VEGA instrument
of the CHARA Array and the instrument CLIMB \citep{climb} for 3T group delay tracking.
The telescopes E1, E2 and W2 were used giving access to ground baselines of 66,
156, and 221m. The seeing was stable with a value of r0 between $10$ and $13$ cm. For
the absolute calibration of the squared visibilities, we observed two different
calibrators, C1 = HD 55185 and C2 = HD 46487. The observing sequence was
C1-T-C2-T-C2-T-C1, each block being almost 20mn. The data were processed using
the standard $V^2$ procedure of the VEGA instrument as described in \cite{vega2}.
To validate the absolute calibration, we first considered C1 as a target and thus
estimated its uniform disk (UD) diameter using C2 as the calibrator. A spectral band of
$\Delta\lambda=30$ nm centered around $\lambda=735$ nm was used for the data
processing. The equivalent UD diameter of C2 is taken as
$\theta_\mathrm{HD46487}=0.180\pm0.013$  mas according to the SearchCal
tool\footnote{Available
at http://www.jmmc.fr/searchcal} of the JMMC \citep{Bonneau06}. Fitting a UD model
to the three squared visibilities provides an estimate of the angular diameter of
HD 55185 of $\theta_\mathrm{HD55185}=0.474\pm0.014$ mas. Spectral bandwidth and
baseline smearing are included but only introduce a bias less than 0.1\% on the
diameter determination. Surface brightness relations, as provided by SearchCal,
provide
an independent estimate of the angular diameter. We found
$0.478$ mas and $0.476$ mas from the $(V-R)$ and $(B-V)$ relations, respectively. Our
result agrees very well with these photometric estimations. We finally use C1 and C2 to calibrate the measurements on the target. To avoid
a strong telluric feature, data are processed using a spectral band of
$\Delta\lambda=20$ nm centered around $\lambda=740$ nm. 
The final error bars in the target measurements take into account the 
measurement noise, the dispersion in the individual measurements, the 
diameter of the calibrators, and their errors.

\section{Determination of the diameter}
To derive a reliable diameter of HD 49933, we base our analysis on a 3D
hydrodynamical simulation of the surface of the star. \cite{allende02} demonstrated that
for Procyon, a star similar to HD 49933, the 3D limb darkening is significantly
different from the corresponding one-dimensional (1D) hydrostatic and homogeneous one. 
The differences were found to be larger in the visible (the VEGA domain) than
  the infrared. We adopt the same methodology as used for both $\alpha$ Cen B \citep{big06} and Procyon \citep{auf05}. A
hydrodynamical simulation of the surface is used to obtain 3D time-dependent limb-darkened intensities, instead of the simplified 1D homogeneous or uniform disk
models, to compute the visibility curves. The angular diameter is found by
minimizing the differences between the synthetic and observed visibility curves.

\subsection{The 3D simulation of the surface}\label{sec:3D}

We use the state-of-the-art radiative hydrodynamical code ({\scriptsize STAGGER
CODE}, Nordlund \& Galsgaard\footnote{1995,
http://www.astro.ku.dk/$\sim$kg/Papers/MHDcode.ps.gz}) to simulate the surface
convection and stratification of HD 49933.  In a
local box,  the code solves the full set of conservative hydrodynamical equations
coupled to an accurate treatment of the radiative transfer.  The code is based
on a sixth order explicit finite difference scheme. The equations are solved on a
staggered mesh where the thermodynamical variables are cell centered, while the
fluxes are shifted to the cell edge. 
 The domain of simulation contains the entropy minimum located at the surface and is
extended deep enough to have a flat entropy profile at the bottom (adiabatic
regime). The code uses periodic boundary conditions horizontally and open boundaries
vertically. At the bottom of the simulation, the inflows have constant entropy and
pressure. The outflows are not constrained and are free to pass through the
boundary.
We used a realistic equation-of-state that accounts for ionization, recombination,
and dissociation \citep{MHD} and continuous + line opacities \citep{gus08}.
Radiative transfer is crucial  since it drives convection through
entropy losses at the surface \citep{stein98}
and is solved using the Feautrier's scheme along several inclined rays (one vertical,
eight inclined) through each grid point. The wavelength dependence of the radiative
transfer is taken into account using a binning scheme in which the monochromatic
lines are collected into 12 bins. The numerical resolution used for the present
simulation is $240^3$. The geometrical sizes are $21\times 21$ Mm horizontally and 5
Mm vertically.
The horizontal sizes of the domain are defined to contain a sufficient number of
granules at each time-step and the vertical one to be deep enough to ensure an
adiabatic regime at the bottom. The stellar parameters that define our 3D model are $\teff = 6690\pm 35$ K, $\logg
= 4.21$ and a scaled solar chemical composition \citep{asplund09} down to $ -0.5$ 
dex. The uncertainty in $\teff$ represents the fluctuations with time around the
mean value. The simulation was run for a couple of stellar hours to get the full
hydrodynamical relaxation. The convection in such an F star is more efficient than
for the Sun: the rms vertical velocity is 3.9 km/s, which is about twice the solar
value.
The granulation pattern is shown in Fig.~\ref{granulation}.

\subsection{Limb darkening and visibility curves}
\begin{figure}[]
\hbox{\includegraphics[height=55mm]{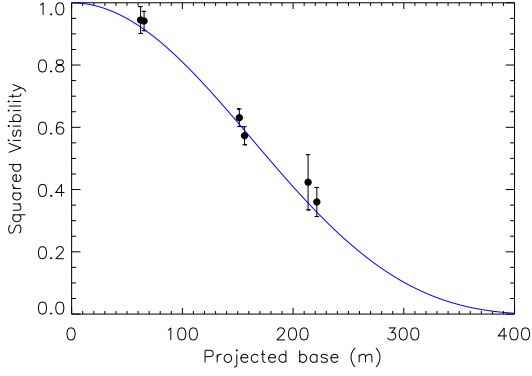}}
\caption{Our best-fit model of the observed squared visibilities (black dots) with the
calculated one (full line) with a reduced $\chi^2=0.47$. The angular diameter is
$\thetaLD= 0.445$ mas.}
\label{V2fit}
\end{figure}

Hereafter, we use the methodology presented in \citet{big06}. The snapshots are taken
after relaxation and cover a sequence of several convective turnover times.
We extract snapshots every five minutes over a sequence of one hour. For each snapshot,
we calculate the radiative transfer in the selected wavelength bandwidth of the VEGA
instrument $\Delta\lambda=[730-750]$ nm to get the monochromatic limb-darkened
intensities for each time step, each position at the stellar surface, and each
cosine $\mu$ of the angle made between the vertical and the line of sight.

The normalized fringe visibilities are obtained using the Cittert-Zernike theorem 
\begin{equation}
V_\lambda(B,\theta) = \frac{\int_0^1 <I_\lambda(\mu)>J_0\left ( \pi B
\theta\lambda^{-1} \sqrt{1-\mu^2}\right ) \mu d\mu}{\int_0^1 <I_\lambda(\mu)> \mu
d\mu},
\end{equation}
where $B$ is the projected baseline (in meters), $J_0$ is the zero order of the
Bessel function, and ${\rm <I_\lambda(\mu)>}$ is the time- and horizontal-average of the
3D intensities. We integrate the monochromatic visibilities over the spectral domain
$\Delta \lambda$ of the instrument weighted by the spectral transmission
$T(\lambda)$
 \begin{equation}
V^2(B,\theta) = \frac{\int_{\Delta\lambda} \left [ V_\lambda(B,\theta)\,
T(\lambda)æ\right]^2 d\lambda}{\int_{\Delta\lambda} \left [ T(\lambda)\right ]^2
d\lambda}.
\end{equation}
The function $T(\lambda)$  is derived from the direct flux measurement of the
target and checked on several bright stars.

\subsection{The $\chi^2$ fit of the data}

The angular diameter is determined by a Levenberg-Marquardt least-squares fit to the
observed squared visibilities with the model. The result is shown in
Fig.~\ref{V2fit}. The value derived from the 3D limb darkening is  $\thetaLD = 0.445
\pm 0.012$ mas. The reduced $\chi^2$ of the fit is 0.47.  Using  the revised
Hipparcos parallax  $\pi=33.69\pm 0.42$ mas  \citep{vanL07}, we obtain a linear
radius of $R=1.42 \pm 0.04 \,R_{\odot}$.
 The corresponding radius derived from 1D hydrostatic limb-darkening
\citep{claret00} is $R= 1.45 \pm 0.04 \,R_{\odot}$, which is significantly larger by 1
$\sigma$ because of the more pronounced limb darkening in 1D. We note that the derived
radius is not very sensitive to the choice of  stellar parameters (those selected in Sect\,\ref{sec:3D}). We indeed tried  several values of $\teff$ (${\rm =
6500,\,6750\,K}$) and $\logg$ (${\rm = 4.0,\,4.5}$) and found very small differences
in terms of derived radius ($\leq 0.002\,R_{\odot}$). 
The derived angular diameter $\thetaLD$  agrees with
the prediction of $0.452\pm 0.007$ mas obtained by the surface brightness relation
\citep{thevenin06}.
The relative uncertainty in the angular diameter ($2.7 \%$) is larger by an order of
magnitude than the smallest uncertainty ever achieved for an interferometric radius,
($0.2\%$) derived in the case of the $\alpha$ Centauri A  \citep{kervella03}.
However, we emphasize that the target in the present work is 20 times smaller. This
uncertainty dominates the total uncertainty of  about $3$\%, including the error in the
parallax.
Nonetheless, the relative uncertainty in the radius is sufficiently
accurate  to tightly constrain the mass using asteroseismic
frequencies \citep{creevey07}.

\section{Asteroseismic fits of acoustic frequencies }

\begin{table*}[t!]
\caption{Our stellar evolution model for HD 49933. The mass ${\rm M}$,  initial
helium content ${\rm Y_0}$, metallicity ${\rm (Z/X)_0}$, core overshoot ${\rm
\alpha_{ov}}$, and mixing length ${\rm \alpha}$ are adjusted to reproduce
the radius obtained by interferometry and the observed large and small separations.}
\begin{tabular}{cccccccccccccc}
\hline
 ${\rm M/M_{\odot}}$  & ${\rm R/R_{\odot}}$ & $\logg$   & ${\rm Y_0}$ & ${\rm
(Z/X)_0}$  & ${\rm \alpha}$ & ${\rm \alpha_{ov}}$ &  ${\rm Age\,(My)}$  &
$\teff\,({\rm K})$  &   $\lum$& ${\rm X_c}$  & ${\rm Y_s}$& ${\rm (Z/X)_s}$ & ${\rm
[Fe/H]}$ \\ 
\hline
     1.200    & 1.42 &    4.21      &   0.29 &   0.016 & 1.00 &  0.35 &  2690 &    
6640 & 0.55  & 0.47 & 0.20 & 0.011 & -0.38 \\ 
\hline
\end{tabular}
\label{model}
\end{table*}

\begin{figure*}[]
\hbox{\includegraphics[height=55mm]{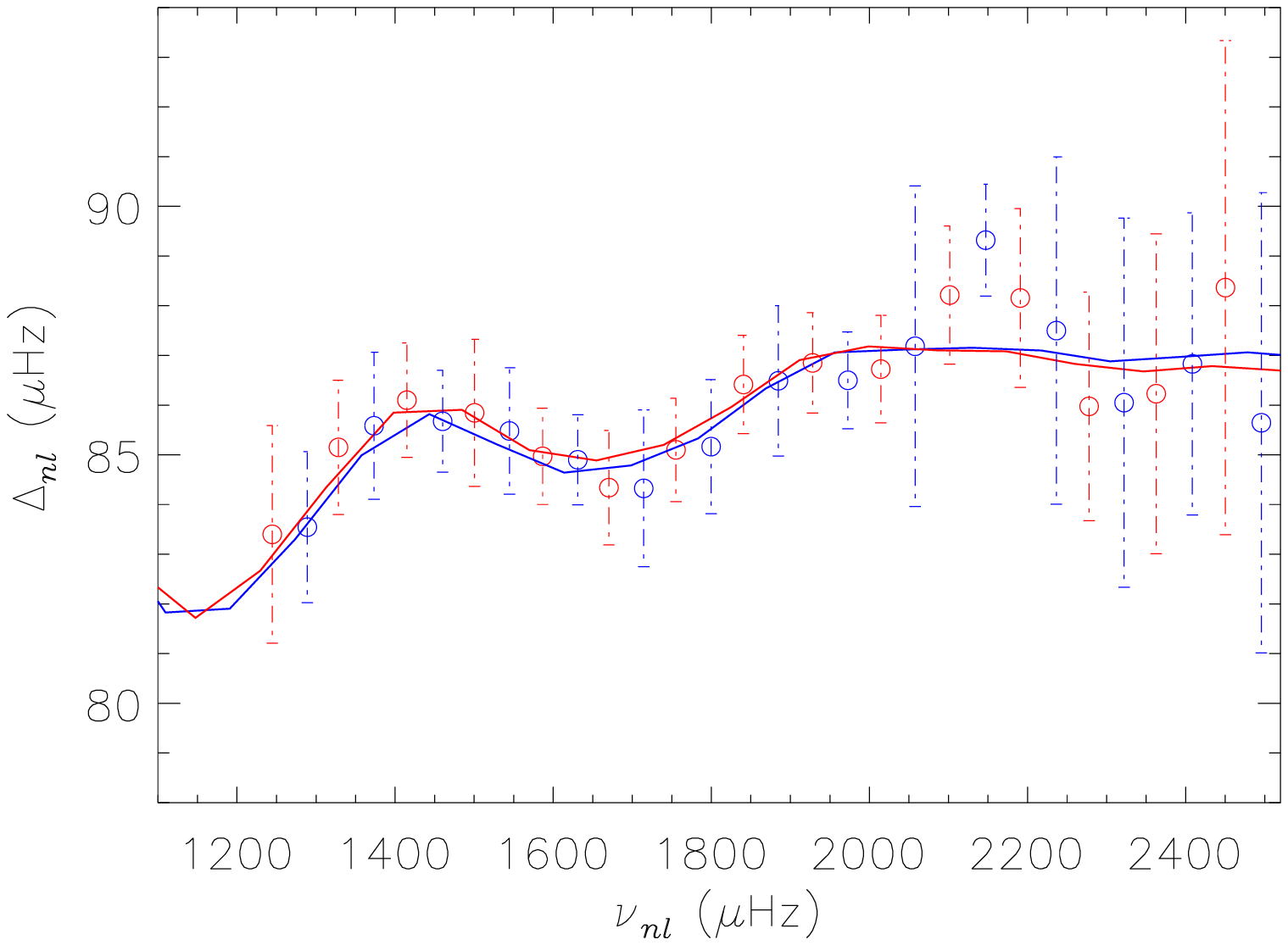}\includegraphics[height=55mm]{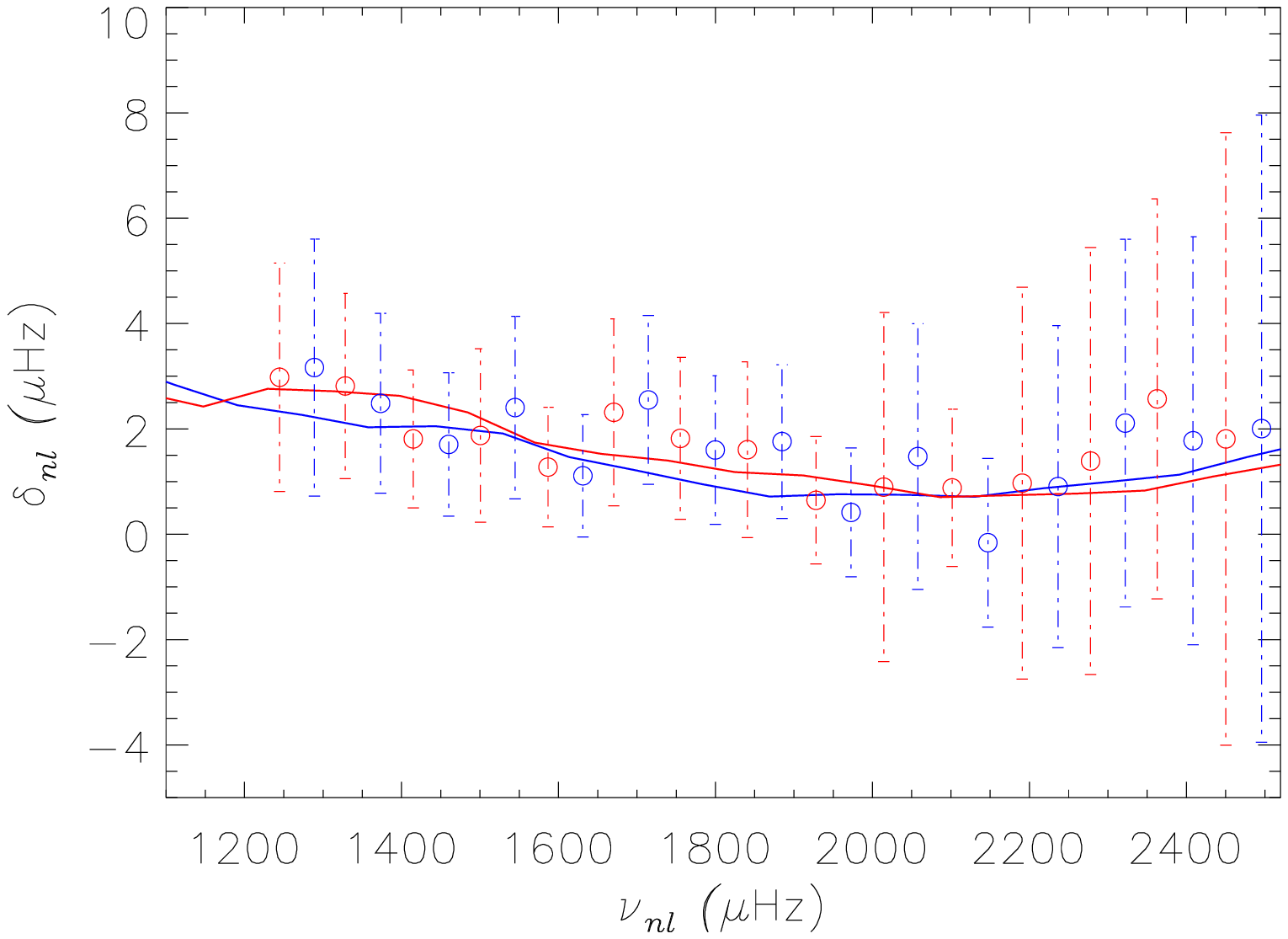}}
\caption{Large (left panel) and small (right panel) separations for
$\ell=0$ (blue) and $\ell=1$ (red) calculated with the stellar model of table \ref{model}. The observed data (bullets) are from
\cite{benomar09}. For each frequency, we took the maximum error bar at $1\sigma$.}
\label{separations}
\end{figure*}

We estimate the mass and the age of HD49933 using
asteroseismic data. 
The aim is to find a stellar evolutionary model whose radius equals the
interferometric one,  whose $\teff$, $\logg$, and ${\rm [Fe/H]}$ match the
spectrometric ones \citep{bruntt09}, and whose  eigenmode frequency separations
match the CoRoT data \citep{benomar09}. 
We compute a grid of models of mass between 1.7 and 2.1 ${\rm M_{\odot}}$, and search for the one that most closely matches the observed small and large frequency separations. 
The stellar model evolves until the radius equals the interferometric one. 
The initial metallicity is adjusted to match the spectroscopic ${\rm [Fe/H]}$. 
The stellar evolutionary models are computed using the {\scriptsize CESAM2k} code
\citep{morel97,morel08}. In addition, we adopt Chemical diffusion \citep{Burgers69} combined with
extra-mixing \citep{morel02} and the \citet{canuto91} formalism of convection. 
The rotation of HD 49933 has a minor impact on the modeling \citep{piau09} and is
ignored. 
For each stellar model, we compute eigenfrequencies between $1100$ and $2500$ $\mu$Hz for
different Legendre polynomial degrees $\ell=0,1,2$ by solving\footnote{Code kindly provided by W. Dziembowski} the standard
linearized equations of continuity and momentum for adiabatic oscillations
\citep[see e.g.][]{bookunno89}. Since these modes pulsate in high overtones ($n$), we added the shift in frequency
caused by the centrifugal distortion of the star \citep{dziem92}.

We fit both the  large and small frequency separations that we define as follows
   \begin{eqnarray}
   \Delta_{n,\ell} =\frac{1}{2} \left ( \nu_{n+1,\ell}-\nu_{n-1,\ell}\right)
\hspace{0.2cm} {\rm and}\hspace{0.2cm}\delta_{n,\ell} =  \nu_{n,\ell} -
\frac{1}{2}\left ( \nu_{n+1,\ell}+\nu_{n,\ell+1}\right).
   \end{eqnarray}
Since the radius is given, the large separations are mainly sensitive to the mass of
the star. The small separations are more sensitive to the deep interior and are
therefore a good indicator of the stellar age \citep[e.g.][]{roxburgh03}.

The solution that minimizes the differences between the observed and computed separations is given in Tab. \ref{model} and the corresponding frequency
separations are shown in Fig.~\ref{separations}. We note that a core overshoot of 
0.35 is  needed to fit the small separations.  We get a better fit of the small separations by including the centrifugal effects on the oscillations.
 The fits to both small and large
separations are very satisfactory and better than in \cite{benomar10}.  To
roughly estimate the errors in the mass and $\teff$, we use the  observed errors in the average
large separation ($\sim M^{1/2}R^{-3/2}$), in the luminosity $L$ ($\sim
R^{2}\teff^4$), and in the interferometric radius.  Our stellar evolution model
agrees well with one of the solutions of \cite{creevey11}, who used a
local minimization method based on radial modes.  The mass  ($1.20 \pm 0.08$ ${\rm
M_{\odot}}$)  is close to that of other works \citep{appourchaux08,benomar10, kallinger10}.
The luminosity ${\rm log}\,L/L_{\odot} = 0.55$ matches the one observed
\citep{michel08}. 
The derived atmospheric parameters ($\teff$ = 6640 $\pm \,100\,${\rm K}, \logg $=
4.21\pm 0.14$, ${\rm  [Fe/H]=-0.38}$)  are in good agreement with the values
obtained by Str\"omgren photometry  \citep{bruntt09}. Our $\teff$
 is larger than the value ($\approx$ 6500 K) derived from ${\rm H_\beta}$
\citep{ryab09,kallinger10}. However, we emphasize that this strong line  is not an excellent
temperature indicator since it  spreads over a wide wavelength range.

\section{Conclusion}
We have reported our long-baseline interferometric observations of the CoRoT target
HD 49933 using the VEGA instrument at the CHARA Array. 
Combined with 3D hydrodynamical simulations of the surface, we derived its radius,
${\rm R = 1.42 \pm 0.04\,R_{\odot}}$, which agrees with the radius derived from the
surface brightness relation.
The use of this interferometric radius reduces the stellar parameter space and
helped us to more tightly constrain the position of the star in the HR diagram. 
The fit of the p-mode frequencies led us to determine the
other stellar parameters. The derived atmospheric ones ($\teff$ = 6640 $\pm
\,100\,${\rm K}, \logg $= 4.21\pm 0.14$, ${\rm  [Fe/H]=-0.38}$) agree well with
photometric studies.

\begin{acknowledgements}
LB thanks A. Chiavassa, R. Collet, and R. Trampedach who provided the latest EOS and
opacities for the 3D modeling, and J. Provost for valuable discussions about the fitting
of frequencies and the stellar evolution modeling.
Computations have been done on the 'Mesocentre SIGAMM'
machine, hosted by the Observatoire de la C\^ote d'Azur.
The CHARA Array is operated with support from the National Science Foundation
through grant AST-0908253, from Georgia State University, the W. M. Keck Foundation,
and the NASA Exoplanet Science Institute.
\end{acknowledgements}

\bibliographystyle{aa}
\bibliography{lbigot}

\end{document}